\documentclass[fleqn,10pt]{wlscirep}
\usepackage[utf8]{inputenc}
\usepackage[T1]{fontenc}
\usepackage{wrapfig}

\newcommand{\revone}{}

\newcommand{\citep}{\cite}

\graphicspath{{figs/},{./},{../}}

\newcommand{\Femp}{F_{\mbox{\scriptsize emp}}}
\newcommand{\Dst}{D_{\mbox{\scriptsize st}}}
\newcommand{{\Krivsky}}{K{\v r}ivsk{\'y}}

\def\ga{\mathrel{\mathchoice {\vcenter{\offinterlineskip\halign{\hfil
 $\displaystyle##$\hfil\cr>\cr\sim\cr}}}
 {\vcenter{\offinterlineskip\halign{\hfil$\textstyle##$\hfil\cr
 >\cr\sim\cr}}}
 {\vcenter{\offinterlineskip\halign{\hfil$\scriptstyle##$\hfil\cr
 >\cr\sim\cr}}}
 {\vcenter{\offinterlineskip\halign{\hfil$\scriptscriptstyle##$\hfil\cr
 >\cr\sim\cr}}}}}

\title{Reconstruction of pretelescopic and early telescopic solar 
activity cycles from auroral records}

\author[1,*]{Kristof Petrovay}
\author[1]{Laura G. Magyar}
\author[2,3,4,5]{Hisashi Hayakawa}
\affil[1]{ELTE E\"otv\"os Lor\'and University, Institute of Physics
and Astronomy, Department of Astronomy, 1117 Budapest,
Hungary}
\affil[2]{Institute for Space--Earth Environmental Research, Nagoya
University, Nagoya, Japan}
\affil[3]{Institute for Advanced Research, Nagoya University, Nagoya,
Japan}
\affil[4]{Space Physics and Operations Division, RAL Space, Science
and Technology Facilities Council, Rutherford Appleton Laboratory,
Harwell Oxford, Didcot, UK}
\affil[5]{Astro-Glaciology Laboratory, Riken Nishina Centre, Wako,
Japan}

\affil[*]{k.petrovay@astro.elte.hu}


\begin{abstract} The historical record of low-latitude aurorae is
essentially a poorly sampled record of the largest space weather
events (SWEs). Its use for the identification of individual solar
cycles is hindered by the low event rate and by the fact that the
solar cycle profile of the occurrence of SWEs does not closely follow
the variation of sunspot numbers. Based on recent studies of the solar
cycle dependence of the occurrence rates of large SWEs, here we
construct  Monte-Carlo simulations of a large number of activity
cycles to identify the optimal procedure to infer the characteristics
of underlying solar cycles from the sparse record. We find that a
reliable reconstruction of the cycle phase ($>90$\% of reconstructed
minima corresponding to actual minima within $\pm 2$ years) is
possible whenever the long-term mean event rate {\revone (annual mean
number of space weather events resulting in low-latitude auroral
sightings)} reaches or exceeds a value around 3. This condition is
found to be satisfied during {\revone most of the} the Early Modern
Active Period (EMAP), a century-long period of normal solar activity
between the Sp\"orer and Maunder Minima. For the numbering of solar
cycles in the EMAP we introduce the ``telescopic era'', where T$n$
denotes the $n$th cycle from the first telescopically observed cycle,
T$0$, ongoing in 1610. Using our optimal procedure we reconstruct a
series of 8 solar activity cycles from T$-5$ to T$2$ (1560--1640).
{\revone Earlier cycles starting from 1540 can be reconstructed
with a somewhat lower degree of reliability. Comparing our results}
with radionuclide-based reconstructions and sunspot observations we
find a good overall correspondence, with the exception of the last
cycle before the Maunder Minimum.   \end{abstract}

\begin{document}

\flushbottom
\maketitle
\thispagestyle{empty}


\section*{Introduction}

Spectacular auroral displays at moderate geomagnetic latitudes have
drawn human attention long before solar flare patrols or geomagnetic
measurements\cite{Vaquero2009}. This resulted in a written record of
space weather activity reaching back several centuries
\cite{KrivskyPejml,Krivsky1996} or even millennia
\cite{Yau1995,Hayakawa:Babylon,Hayakawa2019_assyria,vandersluijs_2023_zhou}.
Beyond providing evidence of individual large space weather events
(SWEs), the auroral record also offers the possibility to study
long-term variations in the Sun's cyclic activity. This possibility
was already recognized in the late nineteenth
century\cite{Fritz:book}. In the following century, much effort was
invested in extending the auroral record with new items and
interpreting it in terms of space climate
variations.\cite{Silverman1992,Vazquez2016}

In addition to secular variations, the possibility of reconstructing
individual Schwabe cycles in the pretelescopic era has also been
explored\cite{arlt2020,hayakawa2024_kepler}.  The first and best
known such reconstruction based on the auroral record was due to
Schove\cite{Schove}.  As this reconstruction was based on the
unjustifiable assumption of long-term phase coherence and it resulted
in predictions that later proved to be wrong, it is now considered of
historical interest only.\cite{Usoskin2023}  Although evidence for an
11-year cyclicity in auroral data has occasionally been
noted\cite{Link1978,Lee2004}, more than three decades have elapsed since the
last attempt\cite{Schroder1992} to pin down the phase and amplitude of
individual historical auroral cycles.

This can be mainly attributed to the success of space climate
reconstructions based on cosmogenic radionuclides.\cite{Usoskin2023}
The availability of increasingly detailed and extensive series of
$^{14}$C and $^{10}$Be abundances from annually stratified samples
(tree rings and ice cores, respectively) resulted in data sets that,
albeit noisy and affected by geographical and climatic factors, have
been considered arguably homogeneous and objective
\cite{beer1998,McBeer2015_annualCR,Usoskin2023}. (Recent studies,
however, noted some caveats.\cite{zheng_2021}) In contrast, the
auroral record consists of sparse, heterogeneous sets of subjective
aurora descriptions by early observers, influenced not only by weather
conditions and a varying geomagnetic dipole but also by political and
cultural factors.\cite{Miyake:book,Stangl2024} Furthermore, even
disregarding these factors, the incidence rate of large geomagnetic
storms and the associated aurorae does not closely follow the sunspot
number curve. Geomagnetic activity normally peaks 1--2 years after the
sunspot number, and this phase difference shows a large scatter from
one cycle to the next.\cite{Vennerstrom2016}  These difficulties led
space climate reconstructions to focus on cosmogenic radionuclide
data.

Radionuclide-based work has by now culminated in a reconstruction of
individual solar cycles covering the last three 
millennia.\cite{Usoskin2021,Usoskin2025,Usoskin2026} It is to be
noted, however, that this $^{14}$C based reconstruction comes with
different quality flags assigned to each cycle, the suggestion being
that for quality flags $\le 3$ (on a scale of 5) cycle parameters or
even cycle identification are uncertain: about half the listed cycles
fall in this category. This is further borne out by a comparison of
different radionuclide-based reconstructions, as discussed later in
this paper.

On the other hand, recent work has elucidated the most salient
features of the geomagnetic storm cycle underlying the auroral cycle
and its relation to the sunspot cycle.\cite{Chapman2020,Owens2021} 
Furthermore, a careful critical compilation of the low-latitude
auroral record was assembled by {\Krivsky}
\cite{KrivskyPejml,Krivsky1996}. A large number of further auroral
sightings have been unearthed in the past three decades, mainly from
East Asia \cite{Yau1995,Lee2004,Hayakawa:Mingshi}. In view of these
promising developments, the problem of reconstructing solar cycles
from the auroral record alone may be worth a second look. 

In the present work, we make a new attempt at reconstructing
individual auroral cycles in the pretelescopic and early telescopic
period based on the existing auroral record. Our reconstruction method
is guided by recent findings\cite{Owens2021} on the cycle profile of
geomagnetic storm incidence and tested in Monte-Carlo 
{\revone (MC)} 
simulations of a
large number of solar cycles following this empirical distribution. 

\begin{figure}[ht]
\begin{center}
\includegraphics[width=0.7\textwidth]{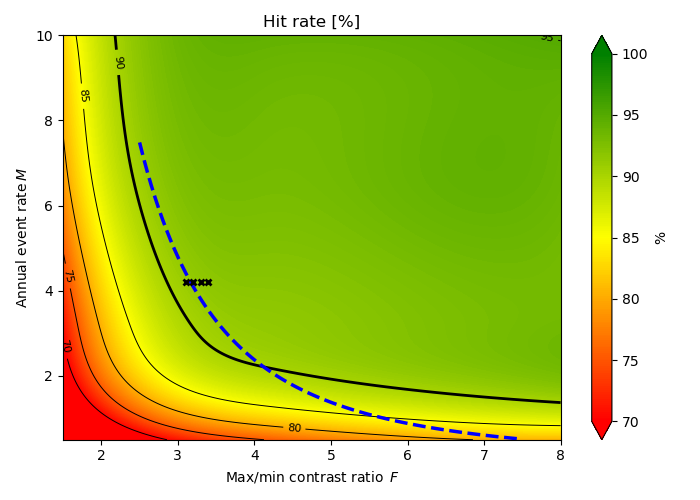}
\end{center}
\caption{Map of the success rate of the identification of cycle minima
against maximum/minimum contrast factor $F$ and mean annual event rate
$M$, based on monte-carlo simulations. The blue dashed line is
representative of the $M$--$F$ relation for the strongest SWE events
at times of normal solar activity. Xs mark possible loci representative of
the EMAP within the $2\sigma$ error interval for the empirically
derived contrast.}
\label{fig:hitmap}
\end{figure}

\section*{Results}

\subsection*{Simulation of auroral cycles}


The cycle variation of the largest geomagnetic storms has been
established in a recent analysis by Owens et al.\cite{Owens2021} (O21
in what follows). The occurrence probability of the storms was found
to be bimodal, split into an active phase and a quiet phase,
represented by a simple step function. The active phase lasts from
solar cycle phase 0.18 to 0.79, while the early and late parts of the
cycle, outside these limits, represent the quiet phase. The occurrence
rate of SWEs is, to a first approximation {\revone (``Phase model'')},
constant in each phase, and a factor $F$ higher in the active phase
compared to the quiet phase, so that the cycle-averaged mean
occurrence rate is $M$ events/year.

In order to test the possibility of reconstructing solar cycles from
auroral records with these characteristics, for a large grid of $F$
and $M$ values, we simulated a series of 1000 cycles following the
above statistical variations, with cycle lengths randomly varying
around 11 years with a standard deviation of 1.15 years, as in the
directly observed sunspot record. In line with usual practice in work
with auroral records, the resulting series of simulated events was
converted to a time series of number of events per calendar year. This
time series was then analyzed by a variety of algorithms to identify
the years of cycle minima. A resulting minimum year within $\pm 2$
years of the actual minimum is considered a successful identification
and the percentage of such successful identifications (hit rate) is
used as a measure of the performance of the method. The procedure
found to perform best essentially consists in smoothing the annual
data by a 12221 slanted boxcar and combining local minima closer than
5.5 years, Further details are given in the Methods section.

{\revone We note that identifying cycle maxima and other notable
epochs was found to be possible with a lower success rate than minima.
This somewhat counter-intuitive conclusion is related to the fact that
the quiet phase is significantly shorter than the active phase; hence,
large storms are scattered randomly over a longer period of time than
the quiet phase between them.}

Fig.~\ref{fig:hitmap} presents the hit rate as a function of $F$ and
$M$. We set a 90\% hit rate (thick contour) as the lower limit for an
acceptable reconstruction. As we may assume that extant records likely
correspond to the most conspicuous aurorae, i.e.\ the largest SWEs,
the mean event rate $M$ in a period of normal solar activity may be
translated to a lower threshold in event size. The statistics in O21
shows that over a period of 150 years, for the top 0.1\% of events
(99.9\% threshold) the mean occurrence rate was $M=0.37$, whereas for
a 99\% threshold it was an order of magnitude higher. On the other
hand, $F$ was also found to depend on the threshold: for the threshold
of 99.9\% O21 quote $F=9$, while from inspection of their figures, for a
threshold of 99\% $F$ is reduced to $\sim 3.5$. Fitting these two
points in the $F$--$M$ plane with a power law results in the dashed
blue curve in Fig.~\ref{fig:hitmap}. Comparing this curve with the
contour line drawn at 90\% we infer that for a reasonably good
reconstruction of solar cycles ($> 90$\% of minima within $\pm 2$
years of the actual date) a mean event rate of $M\ga 3$ is sufficent. 

\begin{figure}[ht]
\begin{center}
\includegraphics[width=\textwidth]{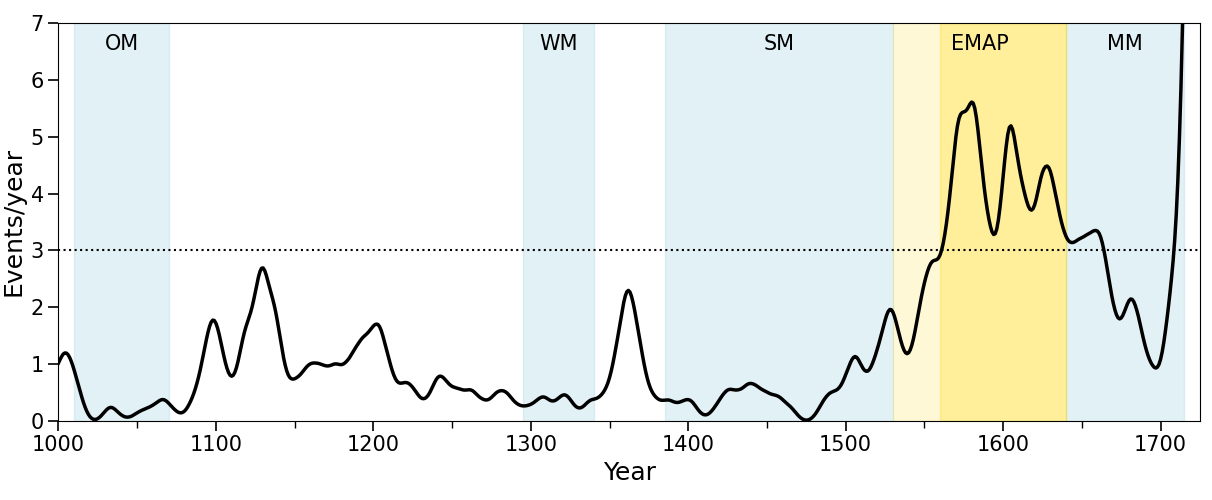}
\end{center}
\caption{Annual record of {\revone SWEs resulting in auroral
sightings} 1000--1700, smoothed with a 25-yr low-pass Butterworth
filter with 2 degrees of freedom. {\revone The conventional
dates\cite{Usoskin2023} of } grand minima of solar activity are marked
with light blue bands: Oort Minimum (OM), Wolf Minimum (WM),
Spörer Minimum (SM), Maunder Minimum (MM). {\revone Light yellow
shading marks the Early Modern Active Period (EMAP), i.e., the
interval between the Sp\"orer and Maunder minima.} 
A dotted horizontal line marks the formal lower threshold for reliable
cycle reconstruction; {\revone the part of EMAP satisfying this condition is
marked with a darker yellow shade.}}
\label{fig:longtermvar}
\end{figure}

\subsection*{Application to the auroral record}

Our starting point for the annual auroral observation record is the
1996 extended version of the {\Krivsky} auroral
catalogue.\cite{KrivskyPejml,Krivsky1996} The variants of this
catalogue have been used to evaluate the long-term solar activity in a
number of previous
studies.\cite{Vaquero:Maunderlimits,Vazquez2016,Chen2021} This
compilation does not simply lump together data from its source
catalogues. Each catalogue item underwent a critical review, and a
significant fraction of reports were discarded as not referring to
aurorae. This critical review was performed by the same authors
uniformly over the whole data set, which lends a certain degree of
homogeneity to this compilation. Such critical reviews have proven
important from some case studies, especially to eliminate possible
contamination.\cite{Usoskin2017,Stephenson_2019,hayakawa_2021_1680,Hayakawa_2026_Wallace}

Beyond the sources used by {\Krivsky}, numerous new historical sources
relating to aurorae have been collected by various authors. Of these,
we collected the records originating before
1700\cite{Yau1995,Hayakawa:Mingshi,Bekli2020,Stangl2021} and merged
them with the {\Krivsky} catalogue. The only exception where pre-1700
records were not used were the large amount of aurora-like sightings
from Korea\cite{Lee2004,Wang2021}. It is currently
believed\cite{He2021,Chen2025}  that a large fraction of these do not
refer to {\it bona fide} polar aurorae but to equatorial airglow
originating in the Western Pacific Anomaly of the geomagnetic field,
affecting this area in the 15th to 18th centuries, so their inclusion
would introduce a strong heterogeneity in the record.

In line with earlier work on auroral data, for the present study we
use annually accumulated records, i.e. the number of {\revone days
with} auroral sightings for each calendrical year. This circumvents
issues related to seasonal variations and allows direct comparison to
the annually sampled cosmogenic radionuclide record. {\revone
Nevertheless, we take into account that large SWEs often last longer
than a calendar day. Therefore, auroral sightings on two consecutive
days are considered by us as a single event. In this sense, the annual
count of events given in our final annual list differs from the number
of days with auroral observation.}

The long-term variation of the record is displayed in
Fig.~\ref{fig:longtermvar}. As is well known, strong inhomogeneities
are present, the most important of which is the order-of-magnitude
increase in the number of auroral sightings from the early 18th
century, due to the start of scientific interest in aurorae. In the
pre-1700 period, dips related to the known grand minima of solar
activity are superimposed on a long-term increase due to the
increasing number of source documents, related, among others, to the
spread of printing. Comparing the long-term mean event rates plotted
in the figure with the $M\ga 3$ lower limit determined in our
simulations, we see that the only extended period where the event rate
reached the required levels was {\revone in} a period of normal solar
activity sandwiched between the Spörer and Maunder Minima. We will
refer to this period as the Early Modern Active Period (EMAP).
{\revone Based on the conventional dates of grand
minima\cite{Usoskin2023}, the EMAP covers the period 1530--1640.
From Fig.~\ref{fig:longtermvar} we see that the period where auroral
cycle minimum epochs can be expected to be reliably determined is
1560--1640, covering a large part of the EMAP.}

\begin{wraptable}{r}{0.5\textwidth}
\begin{center}
\caption{\label{table:aurcycles}
List of reconstructed auroral cycles. The mean event rates per year
$M$ may serve as indicators of cycle amplitude, while the empirical
contrast parameter $\Femp$ can serve as a ``quality flag''. {\revone Cycle
numbering reflects the ``telescopic era'' introduced in this work, where cycle
T0 marks the cycle ongoing in 1610. Reconstructions for the first two cycles,
set in italics, are less reliable.}}
\begin{tabular}{lrrrr}
\hline
&&&\\
Cycle no. & Start & $M$ & Cycle length 
& $\Femp$ \\
&&&\\
\hline
&&&\\
{\sl T-7} & 1540 & 1.7 & 10 & 0.6\\ 
{\sl T-6} & 1550 & 3.0 &  9 & 1.3\\ 
T-5       & 1559 & 2.7 &  7 & 5.4\\ 
T-4       & 1566 & 5.2 & 11 & 3.0\\ 
T-3       & 1577 & 5.5 & 10 & 3.8\\ 
T-2       & 1587 & 3.0 & 10 & 2.1\\ 
T-1       & 1597 & 5.2 & 12 & 2.2\\ 
T0        & 1609 & 3.8 & 10 & 1.4\\ 
T1        & 1619 & 3.9 &  8 & 1.3\\ 
T2        & 1627 & 4.1 & 13 & 1.5\\ 
T3        & 1640 &&\\
&&&\\
\hline
\end{tabular}
\end{center}
\end{wraptable}

All observations in our catalogue originate from geographic latitudes
below $55^\circ$. At the middle of the EMAP this translated to
geomagnetic latitude (MLAT) less than $58^\circ$. The bulk of the
historical sources originate from Central Europe: the MLAT of Prague,
$52$--$53^\circ$ at this time, may be taken as representative. As only
conspicuous aurorae, significantly above the horizon were likely to be
recorded in this early period, the equatorward boundary of the auroral
oval may be estimated to lie at latitudes not higher than $\sim
60^\circ$ during the observed SWEs. This, in turn, translates to
geomagnetic storm amplitudes
$-\Dst>50\,$nT\cite{Yokoyama1998,Niu2015}, meaning moderate or larger
storms. This is consistent with the fact that there are only a very
few years during the EMAP with no auroral observations: indeed, the
annual maximum of $-\Dst$ practically never drops below 50\,nT at
times of normal solar activity\cite{Chapman2020}.

The mean event rate over the {\revone reliable part of the EMAP
(1560--1640) is $M=4.2$.}  The value of $F$ may also be determined
{\it a posteriori} from the reconstructed cycles. This determination,
detailed in the Methods section, yields $3.3\pm 0.2$ ($2\sigma$ error
range) for the underlying contrast parameter. The resulting parameter
combination $(M, F)=(4.2, 3.3)$ is compatible to the pair $(3.7, 3.5)$
corresponding to a 99\% threshold in O21, discussed in the previous
subsection.

\begin{figure}[h]
\begin{center}
\includegraphics[width=\textwidth]{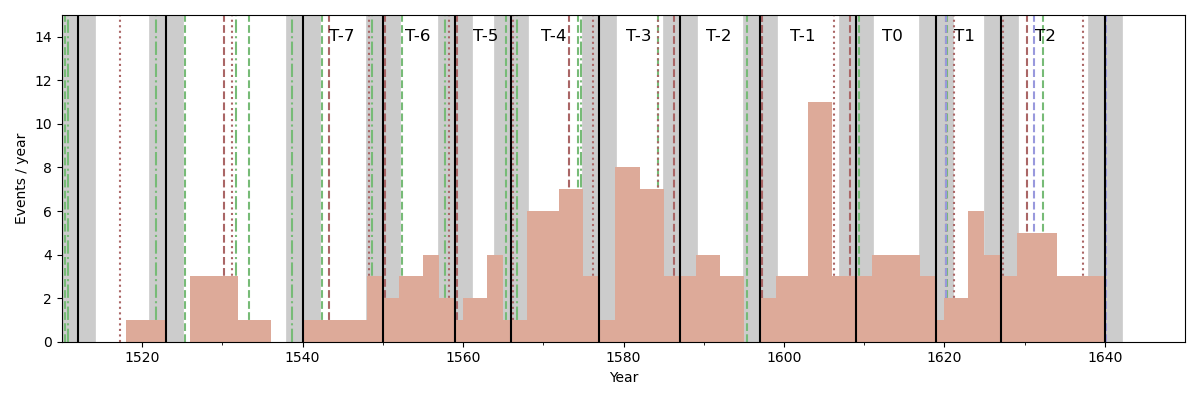}
\end{center}
\caption{Reconstructed auroral cycles in the Early Modern Active Period. Cycle
numbers in our ``telescopic era'' numbering are shown on top. The
bars represent mean number of days with aurorae per year in each of the
unequal cycle quartiles. Solid vertical lines mark the reconstructed minimum
epochs; gray shading indicates the $\pm 2$-year error range. Coloured vertical
lines show minimum years from other reconstructions for comparison. Blue dots:
GSN\cite{Svalgaard+:newGSN}; green dashes: $^{14}$C\cite{Usoskin2021}; green dash-dots:
$^{14}$C\cite{Fogtmann-Schulz2019}; red dots: $^{10}$Be\cite{McB2015}; red dashes:
$^{10}$Be, this work. (Some of the comparison lines were slightly
horizontally shifted for better visibility.)}
\label{fig:auroralcycles}
\end{figure}

\begin{figure}[h]
\begin{center}
\includegraphics[width=\textwidth]{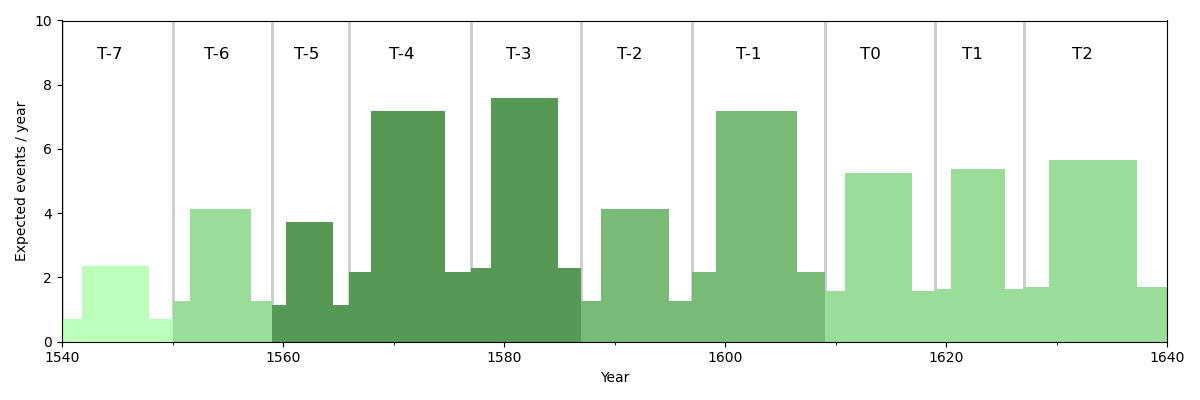}
\end{center}
\caption{SWE probabilities in the Early Modern Active Period based on our
reconstruction results. Cycle numbers in our ``telescopic era'' numbering 
are shown on top. The green bars indicate expected annual event rates in cycle
quartiles (without rounding to calendar years), calculated from the
cycle means given in Table~\ref{table:aurcycles}, assuming the ``Phase'' model
in O21 with the underlying contrast parameter $F=3.3$ deduced for the EMAP.
Shading of the color bars reflects the reliability of the reconstruction as
hinted by the $\Femp$ values in the table. Very light: $\Femp<1$,
light: $1<\Femp<2$; medium: $2<\Femp<3$; dark: $\Femp\ge 3$. }
\label{fig:auroraprobs}
\end{figure}

After these preliminary considerations, we employ our optimized
cycle-finding algorithm to the 1500--1650 section of the annual time
series. The result is shown in Table~\ref{table:aurcycles} and in
Figs.~\ref{fig:auroralcycles} {\revone and \ref{fig:auroraprobs}}. 
Due to the difficulty of identifying cycles during the Maunder Minimum
it is not possible to unambiguously extend the conventional numbering
of solar cycles backwards to the period concerned. Instead, in the
present paper we use the year of the first telescopic sunspot
observations, 1610 as a reference point, denoting the solar cycle
ongoing at this time with T0, while other cycles in the period studied
will be denoted by T$n$, where the integer $n$ is the number of cycles
counted from T0. 

Together with cycle start (phase 0) and mid-cycle (phase 0.5), the
phase values 0.18 and 0.79 separating the quiet and active phases
split each cycle into 4 unequal quartiles, Q1 to Q4. In our annually
discretized data we assign the year of each phase division as the
first year of the new quartile. The annual mean event rate in each of
these quartiles is displayed in Fig.~\ref{fig:auroralcycles} in the
form of bars. Summing up the number of events per quartile for
cycles T$-5$ to T$2$ yields a percentage distribution of events
between the quartiles: 9\%, 39\%, 38\%, 14\% for quartiles 1 to 4,
respectively. Our MC simulations with the same $M, F$ parameter pair
result in: 7\%, 41\%, 39\%, 13\%. (The asymmetry of the distribution
originates from the unequal quartile lengths combined with the
discretization into calendar years.) The close agreement
supports the correctness of the underlying model. 

For the EMAP period 1540--1640 we identify 10 cycles, {\revone the
last 8 of which fall in the period 1560--1640, identified above as
suitable for reliable cycle determination. For these 8 cycles, the
mean cycle length is 10.1 years and the scatter of cycle lengths is
considerably higher than in the recent sunspot record. These
discrepancies may not be statistically significant. Nevertheless, as}
from the MC simulations of the previous subsection the hit rate of our
reconstruction is only expected to be $\sim 90$\%, the possibility
arises that our reconstruction may include a ``fake'' cycle minimum.
The most obvious candidate would be the unusually short cycle T$-5$.
However, this cycle has a high contrast factor, and lumping it
together with any of its two neighbours would result in a cycle with
even more anomalous length. Furthermore, this cycle is also found in
most other reconstructions, as discussed in the next section. Hence,
we do not have any obvious candidate for a misidentified cycle even if
we consider the whole EMAP period.

Prior to cycle T$-7$ our reconstruction is not expected to be trusted
due to the low event rate, and it indeed clearly misses a minimum
around 1531, found by most other reconstructions. 

\subsection*{Comparison with other reconstructions}

Our reconstruction is compared with other solar cycle reconstructions
for this period in Table~\ref{table:mincomparison}. The aurora-based
cycle {\revone reconstructions of Link\cite{Link1978} and
Schr\"oder\cite{Schroder1992} were based on simple visual inspection
of the frequency curve of selected auroral samples.}
Radionuclide-based
reconstructions\cite{McB2015,Usoskin2021,Miyahara2021} involve a heavy
modelling component, including the modulation of cosmic ray flux by
solar activity variations, and carbon-cycle viz. Be-transport models,
respectively. The exception is Fogtmann-Schulz et
al.\cite{Fogtmann-Schulz2019} where the listed epochs were determined
directly from the frequency filtered isotopic abundances. The approach
has the advantage of not being model-dependent, and, while cycle
amplitudes may not be well reproduced, the reconstructed epochs are in
good agreement with other reconstructions. We therefore also applied a
similar approach to the $^{10}$Be isotopic fluxes determined from the
NGRIP ice core\cite{Berggren}, as detailed in the Methods section,
yielding a further set of reconstructed minimum epochs. Finally, the
only reconstruction of the group sunspot number (GSN) overlapping with
the EMAP is also shown.

{\revone When comparing minimum epochs determined by different
recontructions, in each case we identify the 5-year period (true
minimum $\pm 2$ years) that comprises the highest fraction of all
recontructions. The reconstructions falling outside this interval are
then marked as outliers (years set in slanted script in the table).
With only one exception (cycle T2), none of our aurora-based
reconstructions in this work are found to be outliers, supporting the
validity of the method. Nevertheless, the significantly diverging
results concerning the minima of cycles T-5 and T2 indicate that only the
continuous series of 6 cycles from T-4 to T1 can be considered well
established.}

\begin{table}[ht]
\caption{\label{table:mincomparison}
Comparison of reconstructed minimum dates of solar cycles.
The `?' signs were adopted from the original sources of the data. Slanted
typesetting indicates numbers at odds with the other determinations. 
}
\begin{tabular}{lllllllllll}
\hline
&&&\\
Cycle no. & \multicolumn{3}{c}{Auroral minima} & \multicolumn{3}{c}{$^{14}$C based minima}
& \multicolumn{2}{c}{$^{10}$Be based minima} & GSN \\
& This work & L78\cite{Link1978} & Sch92\cite{Schroder1992} &
U21\cite{Usoskin2021} & FS19\cite{Fogtmann-Schulz2019} &
M21\cite{Miyahara2021} & MB\cite{McBeer2015_annualCR} & This work & Sv16\cite{Svalgaard+:newGSN} \\
&&&\\
\hline
&&&\\
T-10 & 1512 &      &       & 1510 & 1511 &      &       & 1509  &  &  \\
T-9  & 1523 &      &       & 1525 & 1522 &      & {\sl 1517}  &       &  &  \\
T-8  &      &      &       & 1533 & 1532 &      & 1531  & 1530  &  &  \\
T-7  & 1540 &      &       & 1542 & 1539 &      &       & 1543  &  &  \\ 
T-6  & 1550 & 1551 & 1552/53  & 1552 & 1549 &           & 1548  & 1550? &  &  \\ 
T-5  & 1559 & 1559 &       &      & 1558 &      & 1558  & 1559  &  &  \\ 
T-4  & 1566 & 1566 & 1566  & 1565 & 1567 &      & 1566? &       &  &  \\ 
T-3  & 1577 & 1576 & 1577/78  & 1574 & 1575 &      & 1577  & 1573  &  &  \\ 
T-2  & 1587 & 1586 & 1589? & 1584 &      &      & 1587  & 1586  &  &  \\ 
T-1  & 1597 & 1601 & 1600/01? & {\sl 1595} &     & 1600 & 1597  & 1597  &  &  \\ 
T0   & 1609 & 1610 & 1610/11  & 1609 &   & {\sl 1606} & 1609  & 1608  &  &  \\ 
T1   & 1619 & 1619 & 1619/20  & 1620 &   & 1621 & 1619  & 1620  & 1620 \\ 
T2   & {\sl 1627} & 1633 &       & 1632 &        & 1633 & {\sl 1627?} & 1630  & 1631 \\ 
T3   & 1640 & {\sl 1645} & {\sl 1636?} &      &      &      & 1637  &  1641 & 1640 \\
&&&\\
\hline
\end{tabular}
\end{table}



\section*{Discussion}

We have established a method to reconstruct solar activity cycles from
the auroral record when the mean annual event rate is at least 3.
Applying this method to the available data a series of 10 cycles have
been reconstructed in the EMAP. 

{\revone We note that the sharp criterion M>3 for the reliability of the
reconstruction is not to be taken very rigorously. First, the scaling relation
plotted in Fig.~\ref{fig:hitmap} on which it is based is itself approximate; the
value of M will depend on the smoothing applied; and the 90\% hit rate required
is also an arbitrary choice. Hence, cycle reconstruction may be attempted also
for periods for somewhat lower values of M. Indeed, as 
Table~\ref{table:mincomparison} shows, our
reconstruction for the period 1540--1560 is also in reasonably good agreement
with the radioisotopic reconstructions.}

It may be worth placing some individual observations in the context of this
reconstruction. The EMAP was a time when auroral reports became increasingly
detailed and objective as aurorae slowly came to be considered a phenomenon of
nature rather than divine portents. The Italian polymath Squarcialupi's
description of the aurora of 10 September 1580, observed from Transylvania, has
been called the first scientific description of aurora
borealis.\cite{Kazmer2016} This intense storm occurred in the 2nd quartile of
the strong cycle T$-3$; this quartile was characterized by the {\revone 2nd}
strongest level of auroral activity in the whole EMAP.

Another recently discussed pretelescopic observation in the EMAP is
Kepler's sunspot of 1607.\cite{hayakawa2024_kepler} The deprojection
of Kepler's drawing allowed to infer that the observed sunspot had low
heliographic latitude, implying that it marked the late phase of cycle
T$-1$. This is consistent with our reconstruction of a minimum date of
1609 for cycle T$0$, in contrast to some other reconstructions
(cf.~Table~2).


\section*{Methods}

The data used in this research and the codes for MC simulations and data
analysis are publicly available on 
GitHub.\footnote{\tt \url{https://github.com/kpetrovay/Auroral-cycles}}
They are accompanied by readme files and the codes are well commented.
In this section we only summarize the most salient technical aspects.

\subsection*{Monte-Carlo approach to optimal cycle identification method}

The main algorithmic steps in the simulation are as follows. 1.
Generate a sequence of 1000 cycles whose lengths are Gaussian random
variables with mean 11 year and $\sigma=1.15$ year. Separate active
and quiet phases in each cycle based on cycle phase. 2. In each time
step generate number of events as a Poisson random variable whose
expected value is computed from $M$ and $F$. 3. Re-discretize time by
calendar years. 4. Analyze the resulting time series of the annual
count of events, evaluating the result in terms of the
success rate, i.e., the fraction of reconstruction of cycle epochs
(minimum, mid-cycle, quiet/active transitions) within $\pm 2$ years,
and select the procedure resulting in the optimal reconstruction.

The tested procedures differed in the following aspects. (a) Smoothing
[frequency filtering with an $n=2$ Butterworth filter, varying
cutoffs; boxcars with different profiles, 121, 12221, 13631, 1222221
etc.] (b) Minimal allowed distance between identified epochs; epochs
closer than this limit were replaced with a weighted average. (c)
Final regularization (epochs separating short-long cycle pairs may be
shifted to reduce difference). 

This resulted in the selection of the following
procedure: (1) 12221 smoothing (2) identify minima and combine minima
separated by $<5.5$ years (3) Regularization: minima between
short--long cycle pairs are shifted by 1 year to improve balance;
short/long defined as deviating min. 3 years from mean. (This step has
very little effect on the final results in practice.)

\begin{figure}[htb]
\begin{center}
\includegraphics[width=0.7\textwidth]{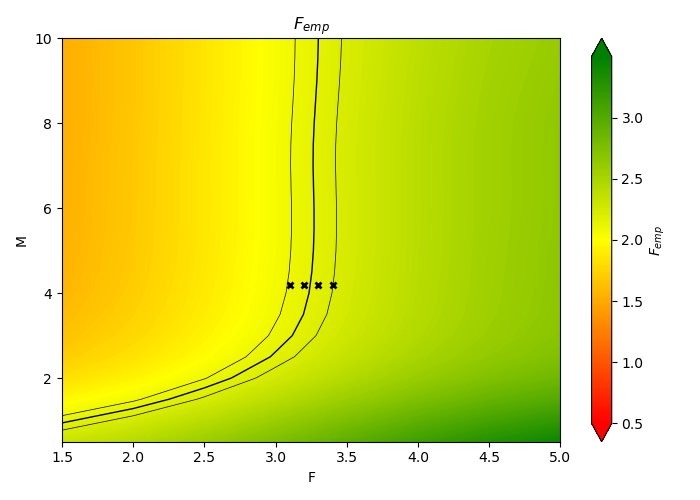}
\end{center}
\caption{Left: Map of the empirically determined contrast factor
$\Femp$ against $F$  and $M$ based on the Monte-Carlo simulations. 
The contour line for $\Femp=2.15$ is drawn in thick, while the
contours for the ${\pm}2\sigma$ limits are drawn in thin. Xs mark
potential loci representative of the EMAP within the statistical
errors.}
\label{fig:Femp}
\end{figure}

\subsection*{Empirical determination of the contrast factor $F$}

An empirical value for the contrast factor may be introduced as
$\Femp=0.64(N_2+N_3)/(N_1+N_2)$, where $N_i$ is the number of events
in the $i$th quartile. Due to the limited length of the data set
({\revone 8 reliable} cycles) this empirical value can only be
determined with a rather large statistical error that may be
determined from the MC simulations. From the historical auroral data
set for the {\revone reliable part of the} EMAP this yields 
$\Femp=2.15{\pm}0.31$.

It needs to be taken into account, however, that due to the annual
discretization, the empirical value falls systematically short of the
actual contrast $F$ of the underlying model. This effect was also
studied in our MC simulated cycles:  Fig.~\ref{fig:Femp} shows $\Femp$
as a function of $F$ and $M$. Together with $M$, for the EMAP this
allows to read off the underlying true contrast value from the plot as
$F=3.3{\pm}0.2$. 

We finally note that our underlying cycle model, corresponding to the
``Phase'' model in O21, might be further improved taking into account
the observation that odd and even numbered cycles tend to have
opposite sign in $N_2-N_3$, i.e. in cycles of one parity storms peak
early, while in cycles of the other parity storms tend to peak late
(``EarlyLate'' model in O21). We performed various tests to check the
presence of such odd-even alternation in the auroral data during the
EMAP but, with the poorly sampled and short data set, no indication of
such behaviour was found. 

\subsection*{Reconstruction of minimum and maximum epochs from $^{10}$Be data}

A proper reconstruction process of solar activity variations from
cosmogenic radionuclide data would involve a comprehensive treatment
to derive the modulation potential of cosmic ray propagation, taking
into account the atmospheric production, residence, and transport of
the nuclide involved, as well as the effects of the time-varying
geomagnetic field\cite{Miyake:book}. While the shorter atmospheric
residence time and higher threshold cosmic ray energy of $^{10}$Be make
this less essential in the case of this nuclide, in the absence of a
proper analysis of the above effects we judge it safer to limit
ourselves to the reconstruction of cycle phases, without
focusing on cycle amplitudes.

{\it Data processing:}

We use the raw $^{10}$Be fluxes FBe10 as determined by Berggren et~al.\cite{Berggren} in
the NGRIP ice core, covering the time period  1389--1995.

Schwabe cycles are not immediately obvious in this series due to short
timescale noise and trends on a longer time scale. To remove these
effects, filtering is applied to the data, retaining only components
varying between time scales $T_1$ and $T_2$. We experimented with
different methods of filtering, including smoothing with a sliding
window (both with sharp cutoff or  a tapered Gaussian
profile\cite{Hathaway:LRSP}) spectral filtering with a sharp cutoff,
and a Butterworth filter. For each filtering method a variety of
values for $T_1$ and $T_2$ were used. The optimal filtering method and
parameters, selected following a calibration process (see below), were
identified as a Butterworth filter of order 2 with $T_1=9\,$yr and
$T_2=18\,$yr. This is in agreement with the original 
analysis\cite{Berggren}. 

The estimated atmospheric residence time of $^{10}$Be is about one year. 
In view of this, we assume that solar activity maxima [minima] precede
minima [maxima] of FBe10 by 1 year.

Local minima  [maxima] in the series of filtered annual flux values
are identifed as imprints of independent solar cycle maxima [minima]
in the record whenever their separation from neighbouring minima
[maxima] exceed $\Delta t=6.5\,$yr. Otherwise, they are considered as
imprints of a single (double peaked) solar maximum [minimum] at an
epoch one year before the weighted average of the indiviual $^{10}$Be
minima [maxima].

{\it Calibration}:

We calibrate our method to the series of directly observed solar
cycles since the end of the Maunder minimum, characterized by the
group sunspot number
(GSN).\footnote{\url{https://www.sidc.be/silso/groupnumberv3}} 
For compatibility with the FBe10 data, annual mean GSN values were
used. The annual GSN series is subjected to the same
filtering method, described above, as the FBe10 values. Comparing our
series of epochs of extrema in the two series, an agreement is
considered to hold whenever an FBe10 minimum [maximum] occurs within
a 5-year interval centered on a date following the GSN maximum
[minimum] by one year. This one-year time shift is introduced to allow
for the atmospheric residence time of $^{10}$Be, as mentioned above.

Note that some further phase shift exists between the sunspot number
and the (inverted) galactic cosmic ray
intensity\cite{Koldobskiy:CRtimelag}. The mean time lag was found to
be 8 Bartels rotations (0.6 years), with a $\pm 100$\% scatter. This
scatter has a bimodal distribution determined by the polarity of the
heliospheric magnetic field (HMF). As this delay is an order of
magnitude shorter than our 5-year time window and information on HMF
polarity is lacking for the period of study in this paper, this
further phase shift is not taken into explicit consideration.

We determine the percentage of such agreements as a percentage of the
total number of minima [maxima] in GSN and in FBe10 for minima and
maxima separately, and consider the agreement best when the average of
these percentage values is highest. This procedure results in the
optimized filtering method and parameters quoted above, where, on
average, the epochs agree in 84\,\%, i.e. an extremum epoch estimated
as the epoch of the extremum of FBe10 plus one year agrees with the
filtered GSN within $\pm 2\,$yrs in 84\,\% of the cases (and to $\pm
1\,$yr in 72\,\%).

Epochs of minima and maxima in the calibration period are listed in
a table available on the GitHub site of this project.

\bibliography{16ctry}

\section*{Funding} 
This research was supported by the NKFIH excellence grants
TKP2021-NKTA-64 and 2026-4.1.1-MISSZIÓK-2026-00005, and by 
the European Union's Horizon 2020 research and
innovation programme under grant agreement no.~955620.  HH received
financial support from JSPS Grants-in-Aids JP25K17436 and JP25H00635,
and the Research Grants in the Humanities of the Mitsubishi Foundation
(202520034).

\section*{Author contributions statement}
K.P.: research concept, auroral data analysis;
L.M.: $^{10}$Be data analysis; H.H.: contribution to discussion 
of early sunspot observations and space weather events.
All authors reviewed the manuscript. 

\section*{Additional information}

\textbf{Competing interests} 
All authors declare that they have no conflicts of interest to
disclose.


\end{document}